\documentclass[aps,prl,showpacs,twocolumn,amsmath,amssymb]{revtex4}
\usepackage{graphicx}
\usepackage{Jonasmacros}
\usepackage{bm}

\begin{document}
\date{\today}
\title{Non-equilibrium triplet blockade in parallel coupled quantum dots}
\author{J. Fransson}
\email{Jonas.Fransson@fysik.uu.se}
\affiliation{Department of Materials Science and Engineering, Royal Institute of Technology (KTH), SE-100 44\ \ Stockholm, Sweden}
\affiliation{Physics Department, Uppsala University, Box 530, SE-751 21\ \ Uppsala, Sweden}

\begin{abstract}
It is theoretically demonstrated that parallel weakly tunnel coupled quantum dots exhibit non-equilibrium blockade regimes caused by a full occupation in the spin triplet state, in analogy to the Pauli spin blockade in serially weakly coupled quantum dots. Charge tends to accumulate in the two-electron triplet for bias voltages that support transitions between the singlet and three-electron states.
\end{abstract}
\maketitle

From fundamental aspects of spin and charge correlations the two-level system in a double quantum dot (DQD) has recently become highly attractive. It has been demonstrated that spin correlations lead to Pauli spin blockade in serially coupled quantum dots (QDs), where the current is suppressed because of spin triplet correlations,\cite{ono2002,rogge2004,johnson2005,franssoncm2005} something which may be applied in spin-qubit readout technologies.\cite{bandyopadhyay2003} Pauli spin blockade has also been reported for general DQDs with more than two electrons.\cite{liu2005} Recently, the Pauli spin blockade with nearly absent singlet-triplet splitting has been employed in studies of hyperfine couplings between electron and nuclear spins.\cite{ono2004,johnson_nature2005,erlingsson2005,koppens2005,petta2005} Besides being present in serially coupled QDs, it is relevant to ask whether an analog of the Pauli spin blockade is obtainable in parallel QDs.

\begin{figure}[b]
\begin{center}
\includegraphics[width=8.5cm]{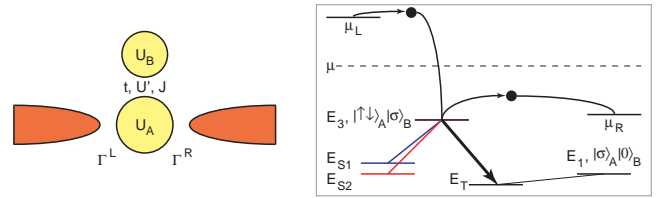}
\end{center}
\caption{(Colour online) Left panel: The coupled QDs of which only one is tunnel coupled to the leads. Right panel: Processes leading to the non-equilibrium triplet blockade. Faint and bold lines signify low and high transition probabilities, respectively. See text for notation.}
\label{fig-system}
\end{figure}

The purpose of this paper is to demonstrate that parallel coupled QDs, see Fig \ref{fig-system}, exhibit regimes of non-equilibrium triplet blockade. Here only one of the QDs is tunnel coupled to the external leads while the second QD functions as a perturbation to the first QD. Important quantities in order to find the non-equilibrium triplet blockade regime is that the QDs are coupled through charge interactions, e.g. interdot Coulomb repulsion and exchange interaction, and weakly through tunnelling. In absence of interdot exchange interaction there may be regimes of usual Coulomb blockade in a finite bias voltage range around equilibrium.

In presence of a sufficiently large ferromagnetic interdot exchange interaction the triplet states $\ket{\sigma}_A\ket{\sigma}_B,\ \sigma=\up,\down$ (one electron in each QD with equal spins) and $[\ket{\up}_A\ket{\down}_B+\ket{\down}_A\ket{\up}_B]/\sqrt{2}$ acquire a lower energy than the lowest two-electron singlet (the singlet states being superpositions of the Fock states $\{[\ket{\up}_A\ket{\down}_B-\ket{\down}_A\ket{\up}_B]/\sqrt{2},\ket{\up\down}_A\ket{0}_B,\ket{0}_A\ket{\up\down}_B\}$). Then, the triplet naturally becomes the equilibrium ground state with a unit occupation probability, provided that the two-electron triplet state has a lower energy than all other states. The triplet persists in being fully occupied for bias voltages smaller than the energy separation between the triplet and singlet states, although transitions between the one-electron states and the singlets may open for conduction. However, for larger bias voltages this low bias triplet blockade is lifted as the transitions between the triplet and the one-electron states become resonant with the lower of the chemical potentials of the leads. At this lifting, the current through the system is mediated via transitions between the two-electron singlets and the one-electron states.

The non-equilibrium triplet blockade regime is entered at bias voltages such that transitions between three-electron states and, at least, one of the singlet states become resonant, see Fig. \ref{fig-system}, while transitions between the three-electron states and the triplet lie out of resonance. At those conditions, an electron can enter the DQD from the lead with the higher chemical potential, through transitions between the singlet and the three-electron states. Transitions from the triplet to the three-electron state are suppressed since the bias is lower than the energy barriers between those states. However, electron tunnelling from the three-electron states in the DQD to the lead with the lower chemical potential are supported through transitions from those states to the triplet, since the tunnelling barrier to this lead is sufficiently low. Thereto, the probability for such transitions are about unity whereas the probability for transitions between the three-electron states and the singlet is at most a half. Finally, charge that end up in the triplet through this process is trapped in this state because of the negligible probability for transitions between the triplet and the one-electron states.

It is noticed that, a finite (ferromagnetic) interdot exchange interaction is not a necessary condition for the existence of a non-equilibrium triplet blockade regime. Nevertheless, a ferromagnetic exchange yields a larger degree of freedom in the variation of the interdot tunnelling and, also, allows a higher temperature.

For quantitative purposes, consider two single level QDs $(\dote{A},\dote{B}$, spin-degenerate) with intradot charging energies $(U_A,U_B)$, which are coupled by interdot charging $(U')$, exchange $(J\geq0)$, and tunnelling $(t)$ interactions. Specifically, the DQD is modelled by\cite{sandalov1995,inoshita2003,cota2005} $\Hamil_{DQD}=\sum_{i=A,B}(\sum_\sigma\dote{i\sigma}\ddagger{i\sigma}\dc{i\sigma}+Un_{i\up}n_{i\down})+(U'-J/2)(n_{A\up}+n_{A\down})(n_{B\up}+n_{B\down})-2J\bfs_A\cdot\bfs_B+\sum_\sigma(t\ddagger{A\sigma}\dc{B\sigma}+H.c.)$, where $\bfs_i=(1/2)\sum_{\sigma\sigma'}\ddagger{i\sigma}\hat{\sigma}_{\sigma\sigma'}\dc{i\sigma'}$, $\sigma,\sigma'=\up,\down$, $i=A,B$, are the spins of the two levels. In analogy with the Pauli spin blockade in serially coupled QDs,\cite{ono2002,franssoncm2005} it is required that the lowest one-electron states, the triplet, and the two lowest singlets are nearly aligned, and that the lowest three-electron states lie below the equilibrium chemical potential $\mu$. Hence, $E_T\approx E_{S1}\approx E_{S2}\approx\min_{n=1}^4\{E_{1n}\}<\min_{n=1}^4\{E_{3n}\}<\mu<E_4$, where $E_T$ and $E_{Sn},\ 1,2$, are eigenenergies for the triplet and the two lowest singlet states, respectively, whereas $E_{1n}$, $E_{3n}$, and $E_4$ are the energies for the one-, three-, and four-electron states, respectively. This requires that $\mu-\dote{B}\approx\Delta\dote{}$, $U'\approx\Delta\dote{}$, and $U_A\approx2\Delta\dote{}\leq U_B$, where $\Delta\dote{}=\dote{B}-\dote{A}$. The inequality $U_A\leq U_B$ points out that the QDs do not have to be identical, merely that the charging energy of the second QD should be lower bounded by the charging energy of the first. It should be emphasized, however, that the presence of the second QD is essential in order to obtain the effect discussed in this paper. Finally, weakly coupled QDs, e.g. $\xi=2t/\Delta\dote{}\ll1$ implies that the energies for the lowest one- and three-electron states acquire their main weight on QD$_A$. This condition yields a low (large) probability for transitions between the triplet and the lowest one-electron (three-electron) states.

In general there are 16 eigenstates of $\Hamil_{DQD}$, labelled $\{\ket{N,n},E_{Nn}\}$ denoting the $n$th state of the $N$-electron ($N=1,\ldots,4$) configuration at the energy $E_{Nn}$.\cite{franssoncm2005} In diagonal form, the DQD is thus described by $\Hamil_{DQD}=\sum_{Nn}E_{Nn}\ket{N,n}\bra{N,n}$. Taking the leads to be free-electron like metals and the (single-electron) tunnelling between the DQD with rate $v_{k\sigma}$, the full system can be written as\cite{franssoncm2005}
\begin{eqnarray}
\lefteqn{
\Hamil=\sum_{k\sigma\in L,R}\leade{k}\cdagger{k}\cc{k}+\Hamil_{DQD}
}
\label{eq-Ham}\\&&
	+\sum_{k\sigma,Nnn'}[v_{k\sigma}(\dc{A\sigma})_{NN+1}^{nn'}
		\cdagger{k}\ket{N,n}\bra{N+1,n'}+H.c.],
\nonumber
\end{eqnarray}
where $(\dc{A\sigma})_{NN+1}^{nn'}=\bra{N,n}\dc{A\sigma}\ket{N+1,n'}$ is the matrix element for the transitions $\ket{N,n}\bra{N+1,n'}$. The operator $\dc{A\sigma}$ signify that electrons tunnel from molecular like orbitals in the DQD through QD$_A$ to the leads, which appropriately describes the physical tunnelling processes.

Following the procedure in Ref. \onlinecite{franssoncm2005}, the occupation of the eigenstates are described by a density matrix $\rho=\{\ket{N,n}\bra{N,n}\}$. In the Markovian approximation (sufficient for stationary processes) one thus derives that the equations for $P_{Nn}\equiv\av{\ket{N,n}\bra{N,n}}$ to the first order in the couplings $\Gamma^{L/R}=2\pi\sum_{k\in L/R}|v_{k\sigma}|^2\delta(\omega-\leade{k})=\Gamma_0/2$ between the DQD and the leads, can be written as
\begin{widetext}
\begin{eqnarray}
\ddt P_{Nn}&=&
\frac{1}{\hbar}\sum_{\alpha=L,R}\biggl(
	\sum_{n'}\Gamma_{N-1n',Nn}^\alpha
		[f^+_\alpha(\Delta_{Nn,N-1n'})P_{N-1n'}
			-f^-_\alpha(\Delta_{Nn,N-1n'})P_{Nn}]
\nonumber\\&&
	-\sum_{n'}\Gamma^\alpha_{Nn,N+1n'}
		[f^+_\alpha(\Delta_{N+1n',Nn})P_{Nn}
			-f^-_\alpha(\Delta_{N+1n',Nn})P_{N+1n'}]\biggr)=0,
\label{eq-dtN}\\
N&=&1,\ldots,4,
\nonumber
\end{eqnarray}
\end{widetext}
where $P_{-1n}=P_{5n}\equiv0$. Here, $\Delta_{N+1n',Nn}=E_{N+1n'}-E_{Nn}$ denote the energies for the transitions $\ket{N,n}\bra{N+1,n'}$, while $\Gamma_{Nn,N+1n'}^{L/R}=\sum_\sigma\Gamma^{L/R}(\dc{A\sigma})_{NN+1}^{nn'}$. Also, $f^+_{L/R}(\omega)=f(\omega-\mu_{L/R})$ is the Fermi function at the chemical potential $\mu_{L/R}$ of the left/right $(L/R)$ lead, and $f^-_{L/R}(\omega)=1-f^+_{L/R}(\omega)$. Effects from off-diagonal occupation numbers $\av{\ket{N,n}\bra{N,n'}}$, which only appear in the second order (and higher) in the couplings, are neglected since these include off-diagonal transition matrix elements to the second order (or higher) which generally are small for $\xi\ll1$.

Since the low bias triplet blockade can be found for weakly coupled QDs whenever $J>0$ is sufficiently large, the following derivation focus on the non-equilibrium blockade. The non-equilibrium blockade discussed here, is driven by opening transitions between the two- and three-electron states. For simplicity, assume that the bias voltage $V=(\mu_L-\mu_R)/e$ is applied such that $\mu_{L/R}=\mu\pm eV/2$. Then for $|eV|<7\Delta\dote{}/4$, $k_BT<0.01U_A$, and $\xi<0.2$, which is sufficient for the present purposes, only the population numbers $P_{1n},\ n=1,2$, $N_T=P_{2n}/3,\ n=1,2,3$, $P_{24},\ P_{25}$, and $P_{3n},\ n=1,2$, are non-negligible. The other populations are negligible since the corresponding transition energies lie out of resonance. Because of spin-degeneracy it is noted that $P_{1n}=N_1/2,\ n=1,2$, and $P_{3n}=N_3/2,\ n=1,2$, which reduces the system to five equations for the population numbers. As discussed in the introduction, the non-equilibrium blockade arises when transitions between a singlet and the three-electrons state are resonant. Therefore, the bias voltage is tuned into the regime where $\mu_L$ lies around these transition energies, e.g.\cite{Delta32} $\min_{nn'}\{\Delta_{3n',2n}\}<\mu_L<\max_{nn'}\{\Delta_{3n',2n}\}$, $n=1,\ldots,5$, $n'=1,2$ (here $eV>0$, the case $eV<0$ follows by symmetry of the system). For such biases it is clear that $f_L^+(\Delta_{2n,1n'})=f_R^-(\Delta_{2n,1n'})=1$, $n=1,\ldots,5$, $n'=1,2$, and that $f_R^+(\Delta_{3n',2n})=0$, $n=1,\ldots,5$, $n'=1,2$. It is also clear that the charge accumulation in the triplet is lifted for biases that supports transitions from the triplet to the three-electron states, hence, the bias voltage has to be such that $f_{L/R}^+(\Delta_{3n',2n})\approx0$, that is $\Delta_{3n',2n}=E_{3n'}-E_T>\mu_L+k_BT$, $n=1,2,3$, $n'=1,2$.  Thus, the equations for the population numbers can be written as
\begin{subequations}
\label{eq-Pneq}
\begin{eqnarray}
N_1&=&\frac{1}{p}N_3=\frac{2/3}{1+2p(\kappa/\beta)^2}N_T
\label{eq-P1neq}\\
P_{2n}&=&\frac{1}{2}\frac{L_n^2
	+\Lambda_n^2p\sum_\alpha f_\alpha^-(\Delta_{31,2n})}
	{L_n^2+\Lambda_n^2f_L^+(\Delta_{31,2n})}N_1,\ n=4,5,
\label{eq-P2nneq}\\
p&=&\sum_{n=4}^5L_n^2
	\frac{\Lambda_n^2f_L^+(\Delta_{31,2n})}
		{L_n^2+\Lambda_n^2f_L^+(\Delta_{31,2n})}
			\biggl\{3\kappa^2+\sum_{\alpha,n=4}^5\Lambda_n^2
\nonumber\\&&\times
	f_\alpha^-(\Delta_{31,2n})\left[1-
				\frac{\Lambda_n^2f_L^+(\Delta_{31,2n})}
					{L_n^2+\Lambda_n^2f_L^+(\Delta_{31,2n})}
												\right]\biggr\}^{-1}.
\label{eq-p}
\end{eqnarray}
\end{subequations}
Here, ($n'=1,2$, $n=4,5$) $\beta^2\equiv\sum_\sigma|(\dc{A\sigma})^{n'1}_{12}|^2=\xi^2/[(1+\sqrt{1+\xi^2})^2+\xi^2]$, $L_n^2\equiv\sum_\sigma|(\dc{A\sigma})_{12}^{n'n}|^2$, $\kappa^2\equiv\sum_\sigma|(\dc{A\sigma})_{23}^{1n'}|^2=(1+\xi^2)/[(1+\sqrt{1+\xi^2})^2+\xi^2]$, and $\Lambda_n^2\equiv\sum_\sigma|(\dc{A\sigma})_{23}^{nn'}|^2$ are the matrix elements for the relevant transitions. The above relations are due to spin-degeneracy, e.g. $\Delta_{2n,11}=\Delta_{2n,12}$ and $\Delta_{31,2n}=\Delta_{32,2n}$, $n=1,\ldots,5$. Using Eq. (\ref{eq-Pneq}), charge conservation ($1=\sum_{Nn}P_{Nn}=N_1+N_T+\sum_nP_{2n}+N_3$) thus implies that
\begin{eqnarray}
N_\text{T}&=&\biggl\{1+\frac{2/3}{1+2p(\kappa/\beta)^2}
	\Bigl(1+p
\nonumber\\&&
		+\frac{1}{2}\sum_{n=4}^5\frac{L_n^2
			+p\Lambda_n^2\sum_\alpha f_\alpha^-(\Delta_{31,2n})}
				{L_n^2+\Lambda_n^2f_L^+(\Delta_{31,2n})}
	\Bigr)\biggr\}^{-1}.
\label{eq-NTneq}
\end{eqnarray}

\begin{figure}[t]
\begin{center}
\includegraphics[width=8.5cm]{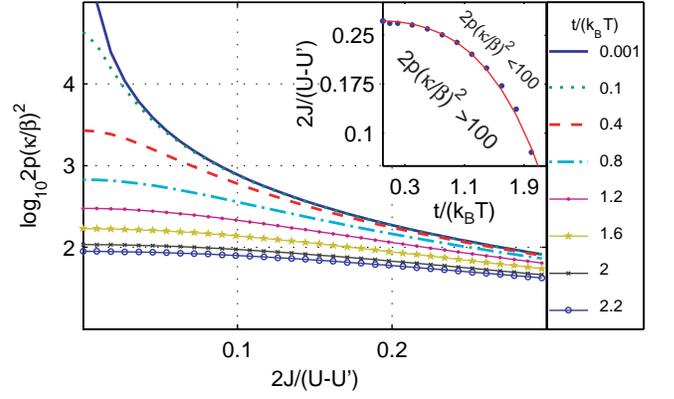}
\end{center}
\caption{(Colour online) Variation of the ratio $2p(\kappa/\beta)^2$ as function of $J$ for different $t$ at constant $\Delta\dote{},\ U',\ U_{A/B}$. The inset shows the region in $(t,J)$-space where $2p(\kappa/\beta)^2>10^2$.}
\label{fig-Jvar}
\end{figure}

Now, the matrix elements $L_n^2,\Lambda_n^2,\ n=4,5$, are finite and bounded, however, $L_4^2,2\Lambda_5^2\rightarrow1$ and $L_5^2,\Lambda_4^2\rightarrow0$ as $\xi\rightarrow0$, hence, the last term in Eq. (\ref{eq-NTneq}) is at most 1/2 for weakly coupled QDs since $p\rightarrow0$, $\xi\rightarrow0$, in the considered bias regime (see discussion below). However, the ratio $2p(\kappa/\beta)^2$ is finite for all $\xi$ and $J>0$, while it diverges as $\xi\rightarrow0$ for $J=0$, see main panel in Fig. \ref{fig-Jvar}. For weakly coupled QDs one thus finds that $N_T\approx1/(1+[1+2p(\kappa/\beta)^2]^{-1})\approx1$ whenever $2p(\kappa/\beta)^2\gg1$. The inset of Fig. \ref{fig-Jvar} illustrates a subset in $(t,J)$-space where this ratio is larger than $10^2$. At this condition, the boundary is approximately given by $J(t)=J_0-15t^2[1+(10t)^2]$.

Using the transport equation derived in Ref. \onlinecite{jauho1994}, identifying $G^<_{Nn,N+1n'}(\omega)=i2\pi P_{N+1n'}\delta(\omega-\Delta_{N+1n',Nn})$ and $G^>_{Nn,N+1n'}(\omega)=-i2\pi P_{Nn}\delta(\omega-\Delta_{N+1n',Nn})$, the current in the considered regime is given by
\begin{eqnarray}
I&=&\frac{e\Gamma_0}{6\hbar}\biggl[3(\beta^2-\kappa^2)
	+\sum_{n=4}^5[L_n^2-\Lambda_n^2f_L^-(\Delta_{31,2n})]
\nonumber\\&&
		+2\sum_{n=4}^5\Lambda_n^2f_L^+(\Delta_{31,2n})
			\frac{L_n^2+p\Lambda_n^2\sum_\alpha f_\alpha^-(\Delta_{31,2n})}
				{L_n^2+\Lambda_n^2f_L^+(\Delta_{31,2n})}\biggr]
\nonumber\\&&\times
			\frac{N_T}{1+2p(\kappa/\beta)^2}.
\label{eq-Jneq}
\end{eqnarray}
This expression clearly shows that a large value of $2p(\kappa/\beta)^2$ yields a suppression of the current, that is, at the formation of a unit occupation in the triplet state. For biases such that $\mu_L<\min_{nn'}\{\Delta_{3n,2n}\}$ is follows that $f_L^+(\Delta_{3n',2n})\approx0\ \Rightarrow\ p\approx0$, which accounts for a lifting of the triplet blockade where the current is $\sim2p(\kappa/\beta)^2$ larger than in the blockaded regime.

The non-equilibrium triplet blockade depends on the interplay between $J$ and $t$. A reduced $t$ leads to a strong localisation of the odd number states in either of the QDs, which for $\Delta\dote{}>0$ leads to that the lowest odd number states are strongly localised on QD$_A$. Then, the probability for transitions between the triplet, and the one-/three-electron states is small/large $(\beta\rightarrow0/\kappa\rightarrow1$).

The singlets, on the other hand, are expanded in terms of the Fock states $\{[\ket{\up}_A\ket{\down}_B-\ket{\down}_A\ket{\up}_B]/\sqrt{2},\ket{\up\down}_A\ket{0}_B,\ket{0}_A\ket{\up\down}_B\}$ with weights that are slowly varying functions of $t$, however, strongly dependent on $J$. A negligible $J$ yields that the two lowest singlets are almost equally weighted on the states $[\ket{\up}_A\ket{\down}_B-\ket{\down}_A\ket{\up}_B]/\sqrt{2}$ and $\ket{\up\down}_A\ket{0}_B$. Increasing $J>0$ redistributes the weights such that the lowest singlet ($\ket{2,4}$) acquires an increasing weight on $\ket{\up\down}_A\ket{0}_B$, whereas the second singlet ($\ket{2,5}$) becomes stronger weighted on $[\ket{\up}_A\ket{\down}_B-\ket{\down}_A\ket{\up}_B]/\sqrt{2}$. Hence, for a finite $J>0$ and $t\rightarrow0$, this redistribution leads to that transitions between the lowest one-electron states and $\ket{2,4}\ (\ket{2,5})$ occur with an enhanced (reduced) probability, e.g.  $L_4^2\rightarrow1,\ (L_5^2\rightarrow0$), and oppositely for transitions between the singlets and the three-electron states, e.g. $\Lambda_4^2\rightarrow0,\ (\Lambda_5^2\rightarrow1/2$). This implies that $p\rightarrow0$ as $t\rightarrow0$ while $p(\kappa/\beta)^2$ remains almost constant. This constant, however, becomes larger (smaller) for smaller (larger) $J$.

\begin{figure}[t]
\begin{center}
\includegraphics[width=8.5cm]{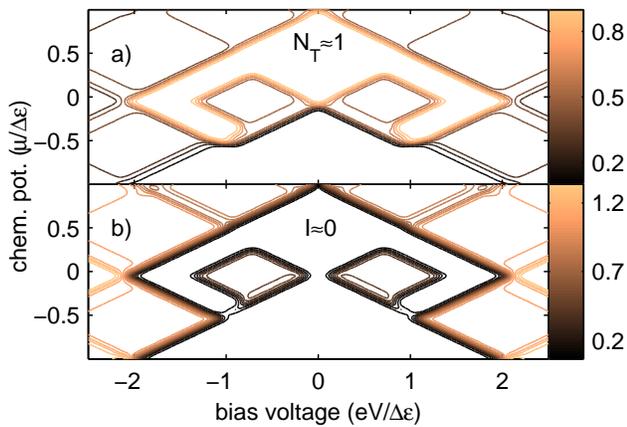}
\end{center}
\caption{(Colour online) Variation of the triplet occupation number $N_\text{T}$ a) and the modulus of the current (units of  $e\Gamma_0/h$) b) as function of the bias voltage and the equilibrium chemical potential $\mu$. Here, $\xi=0.01$, $k_BT=0.01U_A=4t$, and $J=0.2(U_A-U')/2$.}
\label{fig-NT}
\end{figure}
The typical variation of the triplet state occupation number $N_\text{T}$, calculated from Eq. (\ref{eq-dtN}), as function of the bias voltage and the equilibrium chemical potential for $0<J<J_0-15t^2[1+(10t)^2]$ and $t/(k_BT)<2$ is plotted in Fig. \ref{fig-NT} a). Here, varying the equilibrium chemical potential mimics the effect of applying an external gate voltage $V_g$ by means of which the levels of the DQD are shifted relatively the equilibrium chemical potential. The extended diamond marks the region where the occupation of the triplet is nearly unity and where the transport through the DQD is blockaded. The calculated current is displayed in Fig. \ref{fig-NT} b), from which it is legible that the triplet blockade regime is subset of a larger domain of a nearly vanishing current through the DQD. The two diamonds within the low current regime are caused by a lifting of the triplet blockade (see the introduction), where the current is mediated by transitions between the one-electron states and the singlets.

As is seen in Fig. \ref{fig-NT}, shifting $\mu$ in the range $\dote{B}+(\Delta\dote{}-J,2\Delta\dote{})$ causes an extension of the low bias triplet regime since the transitions between the triplet and the one-electron states become resonant at higher biases. On the other hand, the non-equilibrium triplet blockade is shifted to lower biases since $\mu$ lies closer to the transition between the three-electron states and the singlets. The two blockade regimes merge into single as $|\mu-\Delta_{3n',2n}|<|\mu-\Delta_{21,1n'}|$, $n=4,5$, $n'=1,2$, e.g. for $\mu-\dote{B}\in(3\Delta\dote{}/2,2\Delta\dote{})$, see Fig. \ref{fig-NT}. Shifting $\mu$ in the interval $\dote{B}+(\Delta\dote{}/2,\Delta\dote{}-J)$ removes the low bias blockade since the one-particle states become the equilibrium ground state. The non-equilibrium blockade is shifted to lower biases, here caused by transitions between the one- and two-electron states which tend to accumulate the occupation in the triplet.

While the case $\Delta\dote{}>0$ is considered here, the non-equilibrium blockade is also found in the opposite case, e.g. $\Delta\dote{}<0$ and $\mu-\dote{A}\approx\Delta\dote{}$. In this case, however, the system has to be gated such that only the four-electron state lies above $\mu$, whereas the charge accumulation of the triplet state is governed by the same processes as described here.

It should be noted that higher order effects, as well as singlet-triplet relaxation, have been neglected in the equation for the population probabilities $P_{Nn}$. However, in many aspects the situation discussed here corresponds to the experiment reported in Ref. \onlinecite{ono2002}, hence the effect considered should be measurable under much the same conditions. Therefore, as in the case of the serially coupled DQD, the higher order effects give contributions that are at least two orders of magnitude smaller than the second order contributions. Therefore, these can be neglected in the present study. On the same basis as in the description of the serially coupled DQD,\cite{franssoncm2005}  the singlet-triplet relaxation may be neglected here.

The conditions required for the existence of non-equilibrium triplet blockade, concerning the intra- and interdot charge interactions for weakly coupled QDs, have been experimentally obtained for serially coupled QDs.\cite{ono2002,rogge2004,johnson2005} The additional requirement, i.e. a ferromagnetic interdot exchange interaction which is larger than the interdot tunnelling and the thermal excitation energy, is accessible within the present state-of-the-art technology.\cite{kouwenhoven2001,vanbeveren2005,johnson2005,petta2005}

Support from Carl Trygger's Foundation is acknowledged. The Institute of Physics and Deutsche Physikalische Gessellschaft is gratefully acknowledged for covering the publications costs.



\begin{thebibliography}{20}
\bibitem{Delta32} There is one (spin-degenerate) transition energy $(\Delta_{3n,26},\ n=3,4)$ in the vicinity of $\mu$ which is only relevant for $eV\geq U$ since $\min_{n=1}^4\{\Delta_{26,1n}-\mu\}\gtrsim U/2$.
\end{thebibliography}
\end{document}